%% file: galdist_revised.tex
\documentclass[iop]{emulateapj}

\usepackage{times,graphicx}
\DeclareGraphicsExtensions{.pdf,.png}
\usepackage[backref,breaklinks,colorlinks,citecolor=blue]{hyperref}
\usepackage[all]{hypcap}
\usepackage{amssymb,amsmath}
\usepackage{aas_macros}
\usepackage{natbib}
\usepackage{color}
\usepackage{tabularx}
\usepackage{ulem}
\input{defs}

\newcommand{\fbulge}{f_{\mathrm{bulge}}}

\defcitealias{Henderson2014-kmt}{H14}
\defcitealias{KMTref}{Kim et al. 2010}
\input{galdistmacros}

\shorttitle{Is the Galactic bulge devoid of planets?}
\shortauthors{Penny, Henderson \& Clanton}

\begin{document}

\title{Is the Galactic bulge devoid of planets?}

\author{Matthew T. Penny\altaffilmark{1,2}, Calen B. Henderson\altaffilmark{3,4} and Christian Clanton\altaffilmark{5,4}}
\email{penny@astronomy.ohio-state.edu}

\altaffiltext{1}{Department of Astronomy, Ohio State University, 140 West 18th Avenue, Columbus, OH 43210, USA}
\altaffiltext{2}{Sagan Fellow}
\altaffiltext{3}{Jet Propulsion Laboratory, California Institute of Technology, 4800 Oak Grove Drive, Pasadena, CA 91109, USA}
\altaffiltext{4}{NASA Postdoctoral Program Fellow}
\altaffiltext{5}{NASA Ames Research Center, Moffett Field, CA 94035, USA}

\begin{abstract}

Considering a sample of $\nplanets$ exoplanetary systems detected by gravitational microlensing, we investigate whether or not the estimated distances to these systems conform to the Galactic distribution of planets expected from models. 
We derive the expected distribution of distances and relative proper motions from a simulated microlensing survey, correcting for the dominant selection effects that affect the planet detection sensitivity as a function of distance, and compare it to the observed distribution using Anderson-Darling (AD) hypothesis testing. 
Taking the relative abundance of planets in the bulge to that in the disk, $\fbulge$, as a model parameter, we find that our model is only consistent with the observed distribution for $\fbulge<\fbmaxone$ (for a $p$-value threshold of 0.01) implying that the bulge may be devoid of planets relative to the disk.
Allowing for a dependence of planet abundance on metallicity and host mass, or an additional dependence of planet sensitivity on event timescale does not restore consistency for $\fbulge=1$.
We examine the distance estimates of some events in detail, and conclude that some parallax-based distance estimates could be significantly in error.
Only by combining the removal of one problematic event from our sample and the inclusion of strong dependences of planet abundance or detection sensitivity on host mass, metallicity and event timescale are we able to find consistency with the hypothesis that the bulge and disk have equal planet abundance.

\end{abstract}

\section{Introduction}\label{intro}

The planet formation process is complex and the abundance and architectures of planetary systems will likely depend on numerous factors. 
The prime example of such a dependence is the strong correlation of giant planet abundance with host star metallicity~\citep[e.g.,][]{Santos2004,Fischer2005}, and suggests that the formation of giant planets depends on the amount of solids in the protoplanetary disk~\citep{Ida2004}. A correlation of giant planet abundance with host mass~\citep{Johnson2010}, though less secure~\citep[see, e.g.,][]{Lloyd2013}, also points to a similar conclusion, or at least a dependence on total disk mass. 
It is also possible that extrinsic factors such as the presence of a binary companion~\citep[e.g.,][]{Kaib2013}, the density of nearby stars~\citep[e.g.,][]{DeJuanOvelar2012} or the intensity of the ambient radiation field~\citep[e.g.,][]{Thompson2013} may affect planet formation during the protoplanetary disk phase or later by dynamical interactions.
In order to understand the magnitude of each possible effect on planet abundance, it will be necessary to search for planets around stars that span a broad range of fundamental properties and formation environments.

With the advent of modern, high-cadence gravitational microlensing surveys it is now becoming possible to measure the abundance of roughly Neptune-to-Jupiter-mass planets at the orbital distances at which such planets are thought to form (see \citealt{Gaudi2012} for a detailed review).
So far, studies have focused on measuring the overall abundance and mass-function of planets in these orbits~\citep{Gould2010, Sumi2010, Cassan2012, Shvartzvald2016}, but microlensing lightcurves combined with various auxiliary observations can also yield either measurements or constraints on the distance to the planet hosts.
This is particularly interesting for microlensing detections, because microlensing events can be caused by stars at any distance between the Earth and the Galactic bulge, where most of the stars that act as sources reside. 
Therefore, the distributions of host age, metallicity and kinematic population membership (e.g. disk or bulge) will vary as a function of distance, and if the planet abundance is affected by any of these parameters it may be possible to constrain this dependence.
Measuring the relative abundance of planets between the disk and bulge is an especially interesting prospect because it is extremely difficult to probe with any other technique. 
For example, \citet{Clarkson2008} attempted to address this question using 16 planet candidates discovered by the HST Sagittarius Window Eclipsing Extrasolar Planet Survey (SWEEPS), but because of a potentially large false positive fraction they could not draw firm conclusions. 
Our sample of $\nplanets$ microlensing planet hosts is already significantly larger, and future microlensing samples will be larger still and significantly cleaner and easier to model. Additionally, observations from future space-based microlensing surveys will enable the discovery of large samples of transiting planets~\citep{McDonald2014}, whose populations can be determined using similar methods to those of \citet{Clarkson2008}.

Our aim is to investigate what information the distribution of planet host distances inferred by microlensing planet searches might contain about the planet formation process. 
The sample we have to work with is far from ideal for this task. 
It has been constructed through a heterogeneous mix of survey strategies that aim to maximize the number of planet detections, but that do not always allow an easy assessment of the detection sensitivity due to human intervention in the observing schedule and the large, heterogeneous network of follow-up telescopes that are necessary to achieve significant planet sensitivity in many cases.
This has made it extremely difficult to measure the abundance of microlensing planets from microlensing surveys conducted so far: the surveys themselves were until recently not able to observe at high enough cadence to find most planets blind, and follow-up resources needed to be directed in a manner that was independent of the presence of a planet.

These considerations would seem to render our task hopeless with the current data set. 
However, while the presence or not of a potential planetary signature in a microlensing event may affect the way in which it is observed, there is little to no information on the host distance that could affect the decision process of an observer.
This is not to say that the detectability of a planet in a microlensing event is independent of its host distance, just that there is unlikely to be anything that would make the detectability as a function of host distance difficult to model.
This means that if we can construct a reasonable model of both planet detectability as a function of host distance and of the distribution of lens stars (i.e., the potential planet hosts) then we can reasonably compare the distribution of actual planet host distances to that we expect from the model.
More specifically, we can test the hypothesis that the distribution of actual planet host star distances is drawn from the same distribution of host star distances that we produce from our model. 

In this paper we construct a simulated sample of microlensing planet host stars with detected planets and compare the resultant (expected) distribution of host distances with the observed distribution. 
We begin by describing how distances are measured to microlensing planet hosts in \autoref{distmeasure}. 
In \autoref{data} we describe the selection of the observed microlensing planet host sample and in \autoref{model} we describe the model with which we build the comparison distribution. 
In \autoref{results} we present the results of our comparisons of the model and observed distance distributions. 
In \autoref{discussion} we discuss our results and their potential causes, and in \autoref{conclusion} we present our conclusions.

\section{Measuring Distances}\label{distmeasure}

Before discussing the selection of our observed planet host sample it is necessary to first introduce the methods by which lens (host) distances are measured in microlensing events. 
In most single point lens microlensing events the only constraint on the lens distance is provided by the event time scale
\begin{equation}
\tein = \frac{\thetae}{\murelgeo}, 
\label{tein}
\end{equation}
where $\thetae$ is the angular Einstein radius and $\murelgeo$ is the relative proper motion between the lens and source, measured in the geocentric reference frame.\footnote{The geocentric reference frame is an inertial frame moving with the Earth at some significant reference epoch, usually chosen to be the epoch of a prominent lightcurve feature.} 
The angular Einstein radius 
\begin{equation}
\thetae = \sqrt{\kappa M \pirel},
\label{thetae}
\end{equation}
is determined by the lens mass $M$ and the relative lens-source parallax $\pirel$; $\kappa = 8.144$~mas~$\msun^{-1}$ is a constant. 

From an observers standpoint, $\pirel$ is the fundamental quantity related the location of the lens, and is a measure of the proximity of the lens to the source,
\begin{equation}
\pirel = \frac{\text{AU}}{\dl} - \frac{\text{AU}}{\ds},
\end{equation}
where $\dl$ and $\ds$ are the lens and source distances, respectively. 
For most microlensing events, the source lies in the bulge, and it is reasonable to assume that $\ds$ is known to a precision better than $20$~percent. 
This means that $\pirel$ is usually a good indicator of the lens distance. 
Unfortunately, many papers that estimate lens distances do not quote an estimate for $\pirel$ and do not contain enough information to calculate it, so instead we will work with the lens distance estimates $\dl$ to investigate the distribution of lens locations. 

It is possible to measure $\pirel$ and hence obtain a good estimate of the lens distance if two potentially observable quantities are measured: $\thetae$ and the microlensing parallax 
\begin{equation}
\pie = \frac{\pirel}{\thetae}.
\end{equation}
A detailed derivation of microlensing mass and distance measurements using standard notation is given by \citet{Gould2000}, \citet{Gould2013-horne} and \cite{CalchiNovati2016}. 
In all but one planetary microlensing event to date $\thetae$ has been measured from finite source effects~\citep{Witt1994}: these resolve the source star, which provides an angular ruler with which to measure $\thetae$, and hence also $\murelgeo$. 
Measuring $\pie$ is more difficult, and until recently has required detecting subtle distortions to the microlensing event lightcurve that result from the acceleration of the Earth in its orbit~\citep{Gould1992}. 
The chance of detecting these distortions is greatly increased if the event has a long time scale and if the projection of the Einstein ring from the source onto the solar system is small. 
These conditions end up strongly favoring nearby lenses. 
A much less biased way to measure $\pie$ requires observation of a microlensing event from a spacecraft well separated from the Earth~\citep{Refsdal1966,Gould1994} and has recently come to fruition using Spitzer~\citep{Yee2015,Udalski2015-spitzer,CalchiNovati2015,Street2016,Shvartzvald2015-spitzer} and will be further advanced with Campaign 9 of the {\it K2} mission~\citep{Howell2014,Henderson2015}.

If it is not possible to measure the lens distance through measurements of both $\thetae$ and $\pie$, it is possible to estimate the distance through a Bayesian analysis by applying what constraints you happen to have together with priors on the distribution of lenses and sources from a Galactic model, as well as often unstated but implied flat priors such as the abundance of planets as a function of host star mass,  metallicity and orbital architecture. 
In the absence of constraints on the lens distance, these Bayesian estimates will tend to cluster at the peak of the prior probability distribution. 
For this reason, most authors caution against using Bayesian distance estimates for the kind of study we are conducting. 
As there is presently no other way to access distant planets, we choose to commit this sin and throughout remain wary of its consequences (see \autoref{dataprob} for a discussion of its impact).

\section{Data}\label{data}

We select our sample of planet hosts from all planetary microlensing events detected in seasons up to and including 2014 and with a publication date of 2015 or earlier. 
Our only cut is to exclude systems with mass ratios ($q\equiv\mpl/M$) larger than 0.03. 
This cut prevents a strong distance bias that exists for large planets that orbit very low mass stars, such as OGLE-2011-BLG-420 \citep{Choi2013}. If the system has a mass ratio that overlaps that possible for stellar binaries then it is likely, first, that it will not be analyzed in detail, and second, that unless its parallax is measured, it will not be recognized as a planet.

\begin{table*}
\caption{Microlensing planet host stars}
\input{compacttable}
\label{sample}
\end{table*}

\autoref{sample} lists the distances and relative proper motions of the microlensing planet hosts in our sample, and the method by which their distance was measured. Where there is more than one possible solution we use the solution favored by the authors, or where they are agnostic we choose the solution with the lowest $\chi^2$. Often the choice of solution has very little impact of the distance estimate. Where the method column lists $\pie\thetae$, the host distance was \emph{measured} by combining measurements of parallax and finite source effects. Where method is listed as Bayes, $\dl$ was \emph{estimated} using a Bayesian analysis. Many of these Bayesian estimates incorporate constraints from AO imaging.

For two of the events included in our sample, the authors did not compute Bayesian estimates of the lens distance: OGLE-2008-BLG-092~\citep{Poleski2014} and MOA-2013-BLG-220~\citep{Yee2014}. It is important to include these events, because excluding events without parallax measurements would bias the sample toward more nearby lenses. \citet{Poleski2014} argue that their lens is in the bulge due to the small relative proper motion and perform a Monte Carlo simulation to estimate a distance of $8.1$~kpc, which we adopt. \citet{Yee2014} use blended flux constraints and proper motions to argue that their lens resides in the disk, but only provide an approximate distance upper limit of $6.5$~kpc; we adopt a distance of $4.5$~kpc which corresponds approximately to the peak of the disk distance distribution in our model, which is described in the next section. In both cases we arbitrarily adopt an error bar of $2.0$~kpc in order to be able to run a bootstrap analysis in \autoref{discussion}.

\section{Model}\label{model}

For our model we adopt the existing simulation of the KMTNet microlensing survey by \citet{Henderson2014-kmt}, hereafter \citetalias{Henderson2014-kmt}. 
The simulation assumes that the Galaxy follows the \citet{Han2003} model, and tracks whether stars belong to the bulge (which follows a \citealt{Dwek1995} G2 model as described by \citealt{Han1995-tau,Han2003}) or the disk (which follows a \citealt{Bahcall1986} model as described by \citealt{Han1995-pllx}). 
Source and lens stars are drawn from the Galactic model and assigned a weight proportional to their contribution to the event rate, i.e., proportional to $\murelhelio\thetae$. Here, $\murelhelio$ is the relative lens-source proper motion measured in a the heliocentric frame, because the \citetalias{Henderson2014-kmt} simulations do not consider the Earth's orbital motion. Given the Earth's instantaneous velocity of $2\pi$~AU~yr$^{-1}$, the heliocentric and geocentric proper motions can differ significantly, but for the purposes of this paper it is rare for a large relative proper motion in one frame to be small in the other frame. For the rest of this paper then, we will treat heliocentric and geocentric velocities as equivalent when comparing model to data; this is not ideal, but we have no other option as the simulations did not output enough information to reconstruct the geocentric proper motion, and it is not possible to compute the heliocentric proper motion of the observed events unless parallax is measured or the source and lens star are resolved with high-resolution observations long after the microlensing event~(see \citealt{Yee2015} and \citealt{Bennett2007}; \citealt{Bennett2015,Batista2015} for discussions of each option, respectively). We will therefore use $\murel$ to refer to relative proper motions from here on, be they helio- or geo-centric, and assume that their difference does not significantly affect our results.

A planet is assigned to each lens, with mass $M_{\mathrm{p}}$ and semimajor axis $a$ chosen from a grid, and observations of the planetary microlensing event are simulated.
To match observational constraints on the mass function of planets where microlensing is sensitive, we weight each event by a factor $M_{\mathrm{p}}^{0.73}$ \citep{Cassan2012}. 
The simulated data are fit with a single point mass microlensing event, and if the $\chi^2$ of this fit exceeds the $\chi^2$ of the data relative to the underlying planetary microlensing event by 160, the event is flagged as a planet detection. 
Further details of the simulation can be found in \citetalias{Henderson2014-kmt}.

KMTNet plans to observe a small number of fields continuously at high cadence, which is a different observing strategy than was used to find most of the planets in our sample.
Most were found using lower cadence survey observations that covered a larger region of the Galactic bulge, in combination with follow-up observations of individual events, though some of the later discoveries can be classified as survey-only detections.
Without using extremely complicated simulations, we are forced to assume that the observing strategy used will not significantly bias the distance distribution of our sample.
This may or may not be the case, but what is perhaps more likely to affect the distance distribution is the area over which the planets are searched for.
The 4 field survey planned for KMTNet is far smaller than the area over which our sample is distributed. However, \citetalias{Henderson2014-kmt} also simulated a larger area survey at lower cadence in order to optimize the number of fields observed.
We choose to use the widest, 13 field simulation that \citetalias{Henderson2014-kmt} used; this covers a large fraction of the area covered by the OGLE-III survey, with fields between $\ell=7$ and $\ell=-8$ and $b=-2$ to $b=-6$ as well as a couple of fields in the northern bulge. Somewhat surprisingly, we did not find a significant variation of the relative bulge to disk planet host ratio with field location apart from the most outlying southern fields.
With 13 fields, the cadence of observation was $32.5$~minutes.

For this simulation, \citetalias{Henderson2014-kmt} use a relatively sparsely sampled planet mass-semimajor axis grid, with masses of $1$, $10$ and $100\mearth$ and $\log a=0.15, 0.40$ and $0.65$; however, our early work showed that our results were not changed significantly when we used their more densely sampled fiducial simulations covering a smaller area at higher cadence.

As we are interested in the role of metallicity, we assign each lens star a metallicity according to the prescription of \citet{Clanton2014a}. Metallicities for disk stars are drawn from the distributions of [Fe/H] as a function of Galactocentric radius and distance from the plane as measured by APOGEE~\citep{Hayden2014}, and bulge stars are assigned a metallicity drawn from the sample of bulge dwarf metallicities collected by \citet{Bensby2013}.

\section{The Distance Distribution of Planet Hosts}\label{results}

With our sample and model defined we can begin to investigate whether the observed data match our expectations from the model. We can imagine a sophisticated procedure to incorporate all possible information on each planet host and assess the likelihood of a given model in a Bayesian framework. In the future, with more uniform samples of planet hosts this level of detail will be appropriate, but given the non-uniformity of our sample and the difficulty in understanding its selection effects, we elect a simpler frequentist hypothesis testing approach as a first attempt.

\begin{figure}
\includegraphics[width=\columnwidth]{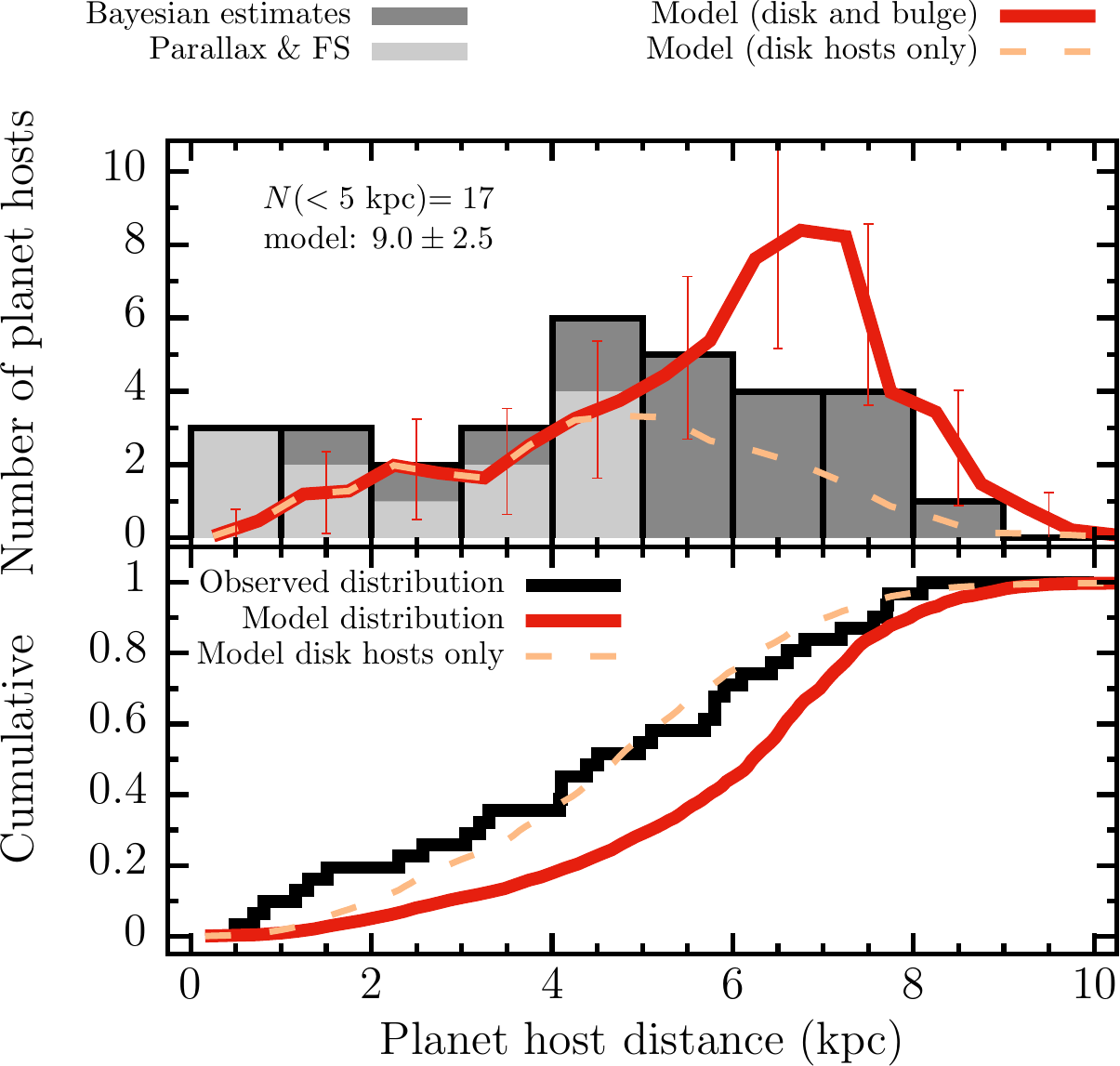}\\
\caption{Histogram and cumulative distribution (black lines) of microlensing planet host distance estimates, shaded by measurement method. Light gray shading represents events with distance \emph{measurements} made by detecting parallax and finite source effects, dark gray represents distance \emph{estimates} usually made using Bayesian techniques. The expected distance distribution from our model is plotted with a red solid line; error bars show the Poisson interval of the model for each bin. The expected contribution from disk hosts is plotted with a dashed orange line.}
\label{distdist}
\end{figure}

The top panel of \autoref{distdist} shows the differential distribution of planet host distance estimates together with the distribution expected from our model. The model (normalized to the total number of hosts) under-predicts the number of hosts in each distance bin less than $6$~kpc, whereas beyond $6$~kpc it generally over-predicts. We have also plotted the contribution of disk hosts and note that the shape of the disk-only host distribution could potentially offer a better fit to the data.

Shown in the lower panel of \autoref{distdist} is the same data in cumulative form. We computed the Anderson-Darling test\footnote{The Anderson-Darling test is similar to, but more sensitive than, the Kolmogorov-Smirnov test.} statistic $A^2=\admodel$ comparing the model\footnote{To convert the discrete model distribution to a continuous distribution, we linearly interpolated the very well sampled cumulative weight distribution.} to the data, which has a $p$-value of $\padmodel$. Comparing only the disk hosts from the model yields $A^2=\addisk$, which has a $p$-value of $\paddisk$. The AD test therefore rejects the hypothesis that the distribution of microlensing planet host distance estimates were drawn from the same distribution as our model, but does not reject the hypothesis that the data were drawn from our simulated distribution including only disk hosts.

\subsection{A Paucity of Bulge Planets?}\label{paucity}

\begin{figure}
\includegraphics[width=\columnwidth]{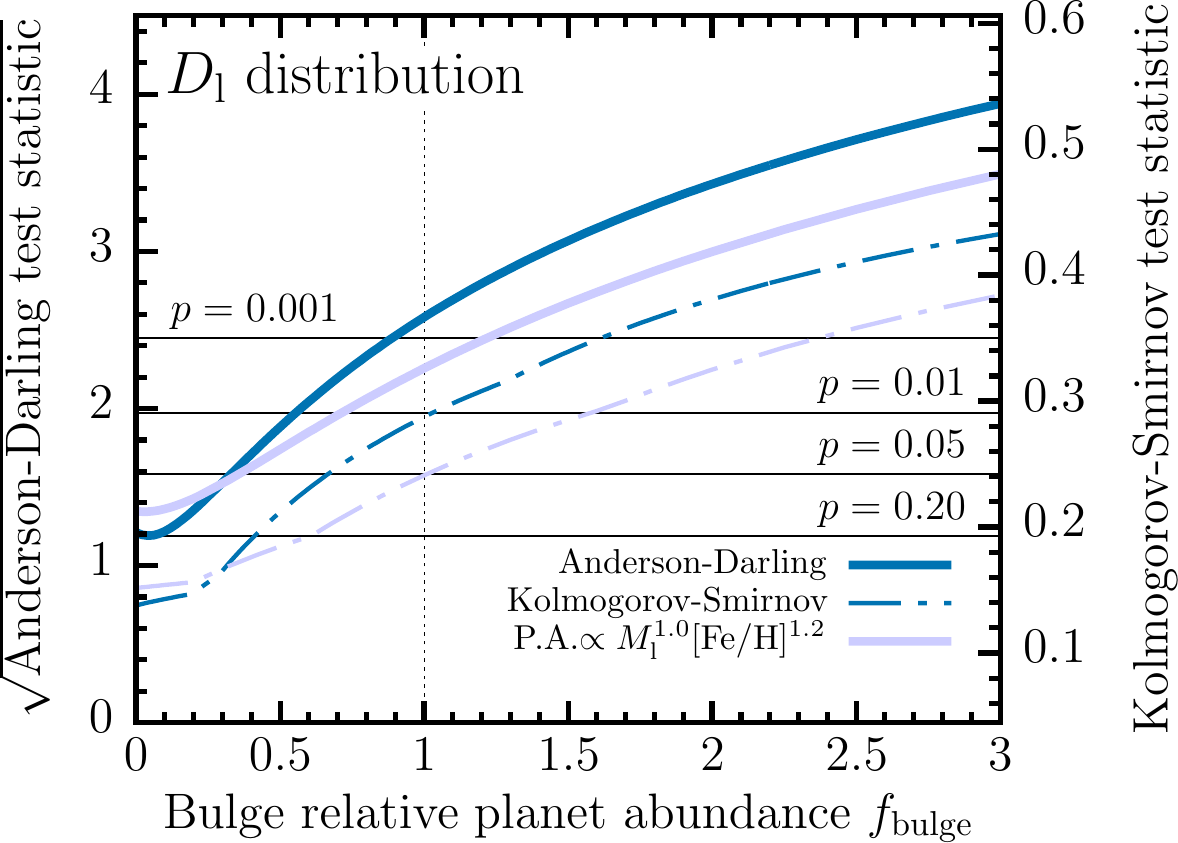}
\caption{The Anderson-Darling (solid) and Kolmogorov-Smirnov (dot dashed, right axis) test statistics plotted as a function of $\fbulge$, the abundance of planets in the bulge relative to the disk. Horizontal lines are plotted at labeled $p$-values for the AD test. The KS test axis is plotted such that there is an approximately linear mapping between AD and KS $p$-values in the $0.001$--$0.2$ range. Lightly shaded lines weight the planet abundance by the joint mass and metallicity correlation found by \citep{Johnson2010}.}
\label{addl}
\end{figure}

Rather than ask the binary question of whether or not there are planets in the bulge, we can ask are there fewer planets in the bulge than the disk? We have weighted the bulge hosts in our model by an additional factor $\fbulge$, the abundance of planets around bulge hosts relative to the abundance around disk hosts and repeated the AD test at various values of $\fbulge$. We plot the results in \autoref{addl} as the dark blue solid line.\footnote{For comparison we have also plotted the less-sensitive KS test; by plotting the square root of the AD test it is possible to plot the range of the KS test statistic such that the $p$-values for each test are mapped approximately one-to-one over the range $p=0.001$ to $0.2$. Note that because the KS test is less sensitive than the AD test, it will typically not reject a hypothesis as strongly and hence give larger $p$-values.} $A^2$ rises as $\fbulge$ increases, and crosses the $p=0.05$ and $p=0.01$ lines at $\fbulge=\fbmax$ and $\fbmaxone$, respectively. The KS test statistic behaves similarly, but crosses the $p=0.01$ line at $\fbulge=0.93$. This result suggests that the abundance of planets in the bulge relative to the disk is less than $\fbmaxone$ at $3$-sigma. Before we can draw this conclusion however, we must more critically examine our assumptions regarding both the data and our model and also reject any other potential explanations.

\subsection{Relative Proper Motions}\label{propmotion}

\begin{figure}
\includegraphics[width=\columnwidth]{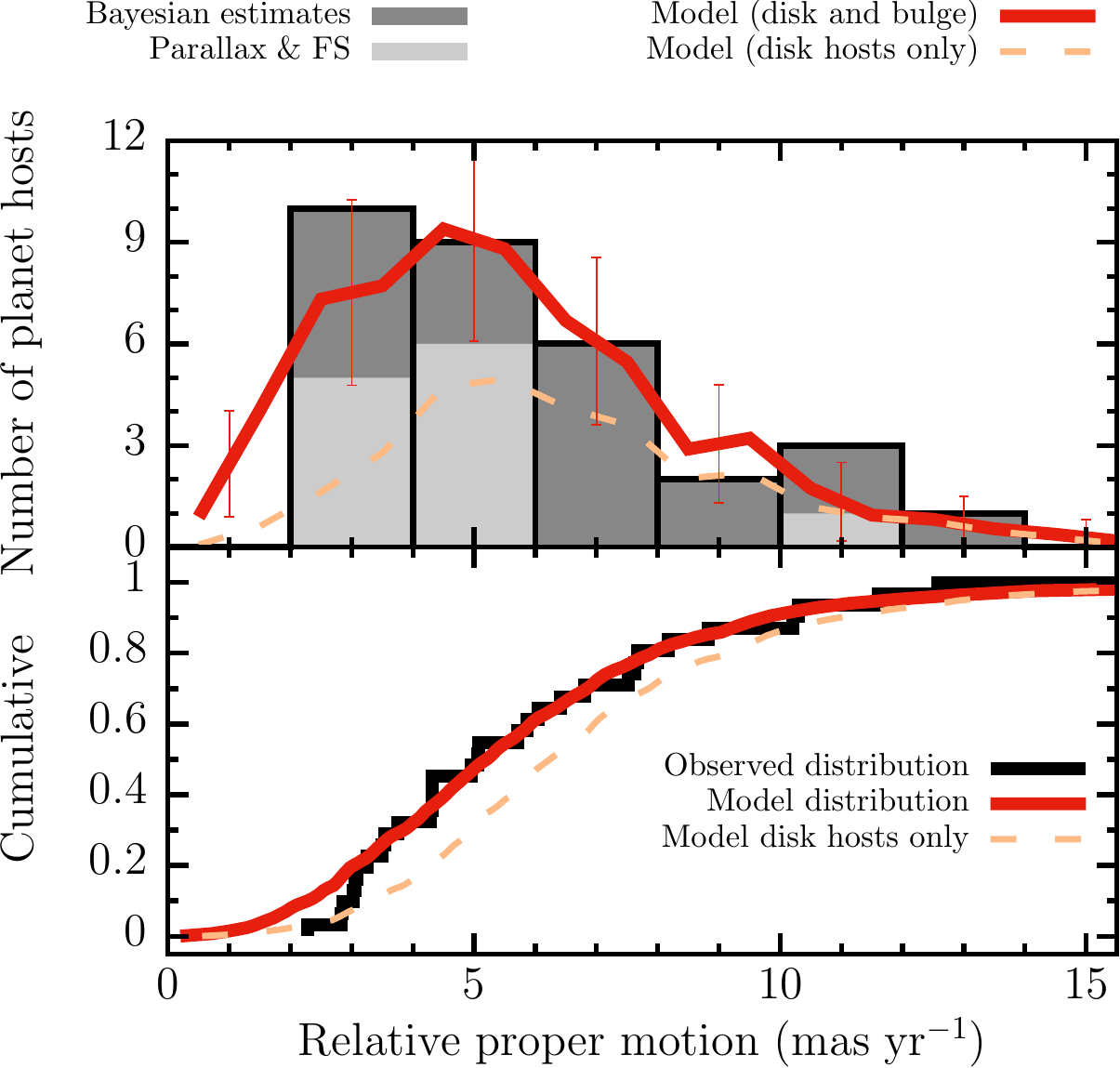}\\
\caption{The distribution of relative proper motions for microlensing events with planets. See \autoref{distdist} for the description of the plot elements.}
\label{mureldist}
\end{figure}

\begin{figure}
\includegraphics[width=\columnwidth]{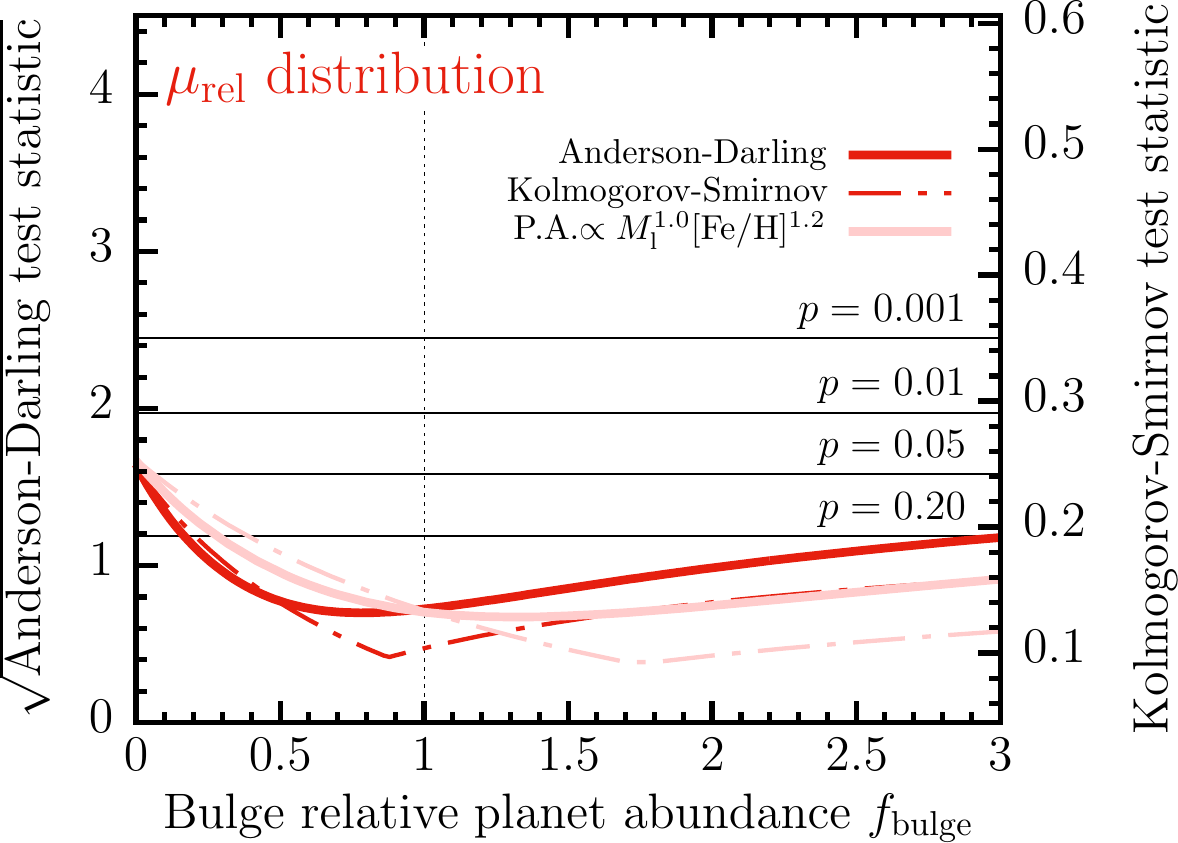}
\caption{AD and KS test statistics as a function of $\fbulge$ for the $\murel$ distribution. See \autoref{addl} for a description of the plot elements.}
\label{admu}
\end{figure}

Given the difficulty in making distance measurements of planet hosts, there are a number of ways that the distribution of distance \emph{estimates} could be altered in a way that we have not modeled. Given that relative proper motions are measured routinely in planetary microlensing events and that stellar kinematics are correlated with population, we can use the distribution of proper motion \emph{measurements} as a consistency check on our findings from the previous subsections. Figures~\ref{mureldist} and \ref{admu} show the results of the same exercise performed with the relative proper motion measurements.

The proper motion distribution should be much more robust to catastrophic errors compared to the distance distribution, but due to the significant overlap of the proper motion distributions of bulge and disk stars, it will have less statistical power to disentangle the contributions of each population. Recall also that we are not comparing relative proper motions in the same frame, so it is possible that we are biasing the result.

The lack of statistical power is borne out, with a large range of $\fbulge$ producing AD test statistics with large $p$-values. However, the AD statistic does rise toward smaller values of $\fbulge$ reaching $\addiskmu$ at $\fbulge=0$, which corresponds to $p=\paddiskmu$ indicating tension with the hypothesis of \emph{no} bulge planets. However, the proper motion distribution does not rule out all of the space with $\fbulge<\fbmaxone$ that is consistent with the distance distribution.

\subsection{Other Physical Explanations?}\label{otherfactors}

\begin{figure}
\includegraphics[width=\columnwidth]{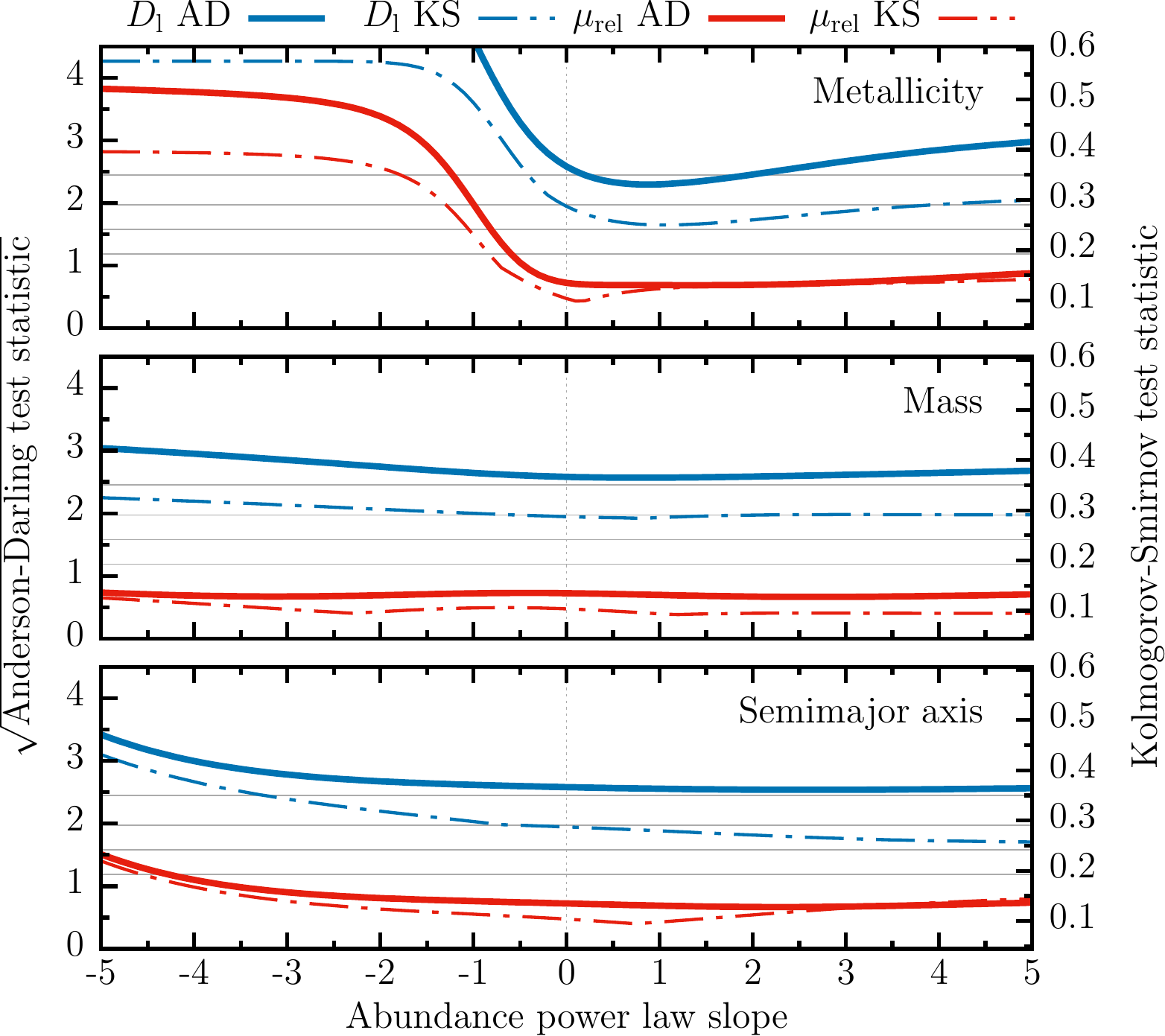}
\caption{AD and KS test statistics as a function of power law slope parameters for toy models (\autoref{toymodels}) of the planet abundance as a function of metallicity (top) host mass (middle) and semimajor axis (bottom) for both the distance estimate $\dl$ (blue) and relative proper motion $\murel$ (red) distributions.}
\label{amassfeh}
\end{figure}

To investigate if other physical factors could cause the observed distribution of distance estimates, we considered toy models of the planet abundance $f$ as a function of host metallicity [Fe/H] , host mass $M$ and planet semimajor axis $a$:
\begin{equation}
f\propto 10^{\alpha\text{[Fe/H]}}, \quad f\propto M^{\beta}, \quad f\propto a^{\gamma},
\label{toymodels}
\end{equation}
 where
\begin{equation}
f = \frac{\dd^3N}{\dd\text{[Fe/H]}\dd M\dd a},
\label{fdef}
\end{equation}
following a long tradition of using power laws to model the distribution of exoplanet abundances \citep[e.g.,][]{Stepinski2000,Lineweaver2003,Fischer2005,Johnson2010}.
Figure~\ref{amassfeh} plots the AD and KS test statistics as a function of each parameter $\alpha, \beta$ and $\gamma$. If a value of $\alpha$, $\beta$ or $\gamma$ different from our fiducial value of $0$ significantly reduces the value of the AD test statistic, then this would indicate that rather than a different abundance of planets in the disk and bulge, a dependence of planet abundance on intrinsic stellar properties or system architecture may be responsible for the discrepancy that we see between the model and the observational data. 

A value of $\alpha=\minal$ improves the model's match to the distance estimate distribution, but only to a value of $A^2=\minalad$ with $p=\pminalad$. Interestingly, negative values of $\alpha\le \alminmu$ are rejected by the AD test with $p\le 0.01$ for the $\murel$ distribution as well as the $\dl$ distribution. Were there a strong negative correlation of planet abundance with metallicity, the fraction of stars in the low-metallicity tail of the bulge's broad metallicity distribution would vastly overproduce planets compared to the disk to the extent that it would even affect the relative proper motion distribution of planetary microlensing events.

Allowing non-zero values of either $\beta$ and $\gamma$ does not significantly reduce $A^2$ (though implausibly strong negative correlations do slightly increase $A^2$), implying that these parameters do not significantly affect the distance estimate distribution. Note however, for both these parameters our model has some shortcomings. First, for both the disk and bulge \citetalias{Henderson2014-kmt} do not simulate events with host stars more massive than $1\msun$. This cutoff is reasonable for the older bulge, where stars more massive would have evolved off the main sequence, but for the younger disk more massive main sequence stars will still be present. If there were a dependence of planet abundance on host mass, the relatively longer lever arm of the disk mass distribution could have an effect on the distance distribution. Second, \citetalias{Henderson2014-kmt} only explore a coarse grid of semimajor axis, so it is not clear if a smooth distribution of semimajor axis would change the result. Because the Einstein ring radius $\re$ depends on $\dl$ and microlensing's sensitivity to planets is strongly peaked around separations of $1\re$, a dependence of planet abundance on semimajor axis would affect the planet host distance distribution.

Given that radial velocity surveys have found a correlation between both host mass and metallicity, we can ask how such a correlation would change our findings from \autoref{paucity}. With no correlation with mass or metallicity ($\alpha=\beta=0$), we found $95$~percent limits of $\fbmin\le\fbulge\le\fbmax$ from AD tests on the $\dl$ and $\murel$ distributions. Weighting by the \citet{Johnson2010} mass-metallicity-planet abundance trend ($\alpha=1.2, \beta=1.0$) produces a qualitatively similar model distribution to our fiducial model, but the number of disk hosts at distances of $3$--$5$~kpc is enhanced at the expense of bulge planets between $6$ and $9$~kpc. This enhancement of planets in the inner disk is due to the APOGEE radial metallicity gradient rising above the mean metallicity of the bulge in this region. The AD and KS statistics as a function of $\fbulge$ including the mass and metallicity correlations are plotted in Figures~\ref{addl} and \ref{admu} with lighter shades. The $95$~percent confidence limits in this case are $\jfbmin\le\fbulge\le\jfbmax$. The 1~percent upper limits on $\fbulge$ increase from $\fbmaxone$ in the no-correlation case to $\jfbmaxone$ with the observed correlations. Note that in the two scenarios, $\fbulge$ has a slightly different interpretation; with no mass or metallicity correlations, $\fbulge$ can be interpreted as the ratio of planets per star in the bulge relative to that in the disk, but with the correlations, $\fbulge$ is only the contribution to this ratio that has not been accounted for already by mass and metallicity correlations.

\section{Discussion}\label{discussion}

So far, while acknowledging potential problems with the distance estimates, we have treated them as if they were correct and investigated the consequences. Now the time has come to examine the distance estimates more critically and ask if they themselves are causing the apparent lack of planets in the bulge. We must also examine the validity of our methods and the accuracy of our model.

An obvious criticism of our work so far is that we have not accounted for the uncertainties in the distance estimates (or proper motions) when computing $p$-values. If we were to do so, the $p$-values would increase. To compute accurate $p$-values we would need to bootstrap resample from the posterior probability distributions of each distance estimate, but these distributions are rarely made available with their papers and are sometimes not even plotted. As an inferior substitute to the posteriors (especially when some are bimodal), we assumed that the quoted uncertainties on each measurement were Gaussian (with asymmetric error bars being treated as two half Gaussians with equal probability), and resampled the $\dl$ distribution $10^4$ times with $\fbulge=1$; distances that were drawn as negative were redrawn until they were positive. $95$~percent of the resamplings had a $p$-value less than $\presample$, compared to the $p$-value we measured for the actual sample of $\padmodel$; and only $\ngepone$ of the resamplings had $p>0.01$. This strongly suggests, but is not conclusive given our simplistic handling, that proper consideration of the posterior distributions would not explain the apparent lack of planets in the bulge.

\subsection{Problems with the model?}\label{modprob}

It is certain that our model is inaccurate at some level. The \citet{Han2003} Galactic model used by \citetalias{Henderson2014-kmt} is relatively old by now in comparison to many new results on the structure and kinematics of the bulge and disk over the intervening time~\citep[e.g.,][]{Clarkson2008,Nataf2010,McWilliam2010,Wegg2013}. Additionally, the disk of \citet{Han2003} does not have a central hole, so one might consider disk stars near the Galactic center to belong to the bulge. But the effect of a lack of a disk hole might be cancelled by not also accounting for stars that were born in the disk and later swept up by the bar. Despite these problems, it is difficult to imagine how the various inaccuracies in the model could combine to form an at least factor-of-three error \citep[for our 95~percent confidence limits accounting for the][mass and metallicity correlations]{Johnson2010} in the \emph{relative} microlensing event rates between the disk and bulge that would be necessary to explain the disagreement between model and data. 

\begin{figure}
\includegraphics[width=\columnwidth]{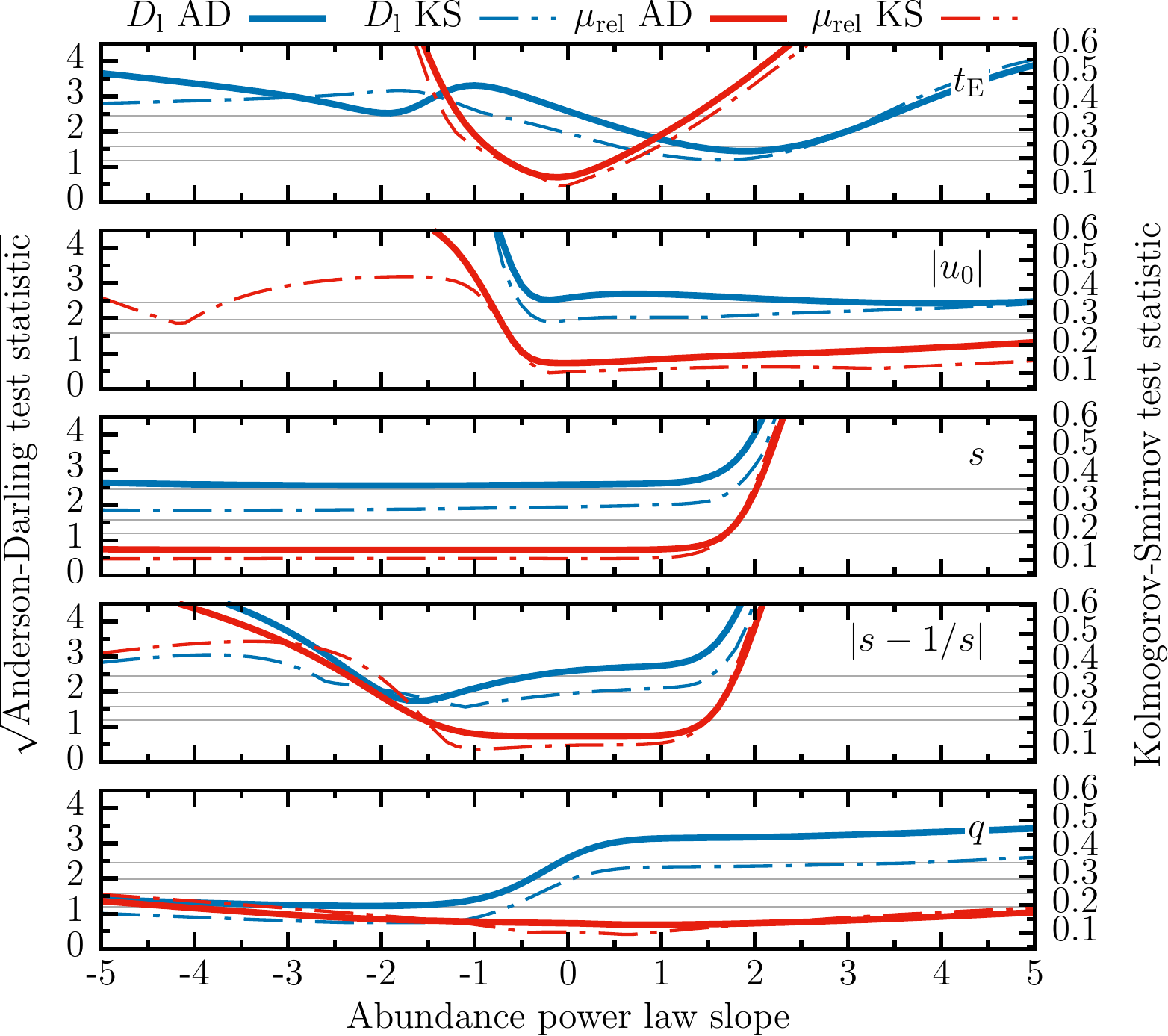}
\caption{As \autoref{amassfeh}, but for microlensing parameters that could affect planet detectability.}
\label{tEsqu0}
\end{figure}

It is possible that the \citetalias{Henderson2014-kmt} simulations do not accurately reflect the process by which most of the planets were found, and therefore alter the distribution of distances. The simplest way such an error may manifest itself is by an additional dependence of planet detection efficiency on the parameters of the microlensing events. We can again use toy power-law models to test how an unaccounted for term in the detection efficiency behaves. The event timescale $\tein$ and impact parameter $\uzero$ are the most likely culprits, but we also test $s$ and $q$, and a derived parameter $|s-1/s|$, which describes the approximate position of the planetary caustics and has a minimum at $s=1$ where microlensing is most sensitive; the results are shown in \autoref{tEsqu0}. 

An additional weighting of $\tein^{{\sim}1}$ improves the match for the $\dl$ distribution ($p=0.023$), but at the cost of increased tension with the $\murel$ distribution ($p=0.013$), and would not completely solve the problem. Such a weighting is a potentially plausible explanation for the lack of planets at large $\dl$ because bulge-bulge lensing events tend to have shorter timescales, but longer timescales allow for more time to organize follow-up observations of high-magnification events, which the \citetalias{Henderson2014-kmt} survey-only simulations would not capture. Applying a $\tein^{+1}$ weighting also brings the median of the model event timescale distribution to $39.5$~d (instead of $23.5$~d for the unweighted distribution) which is more in line with the median timescale of the sample, which is $40$~d.

Despite high-magnification events being over represented in the observed sample relative to the simulations, there is no improvement offered by an inverse scaling with the impact parameter $|\uzero|$. A power law in $s$ does not provide any improvement. A strong negative power of $|s-1/s|^{-1\rightarrow -2}$ does provide an improvement, which would imply that there was a selection effect that favored planets closer to the Einstein ring. This might be a plausible explanation if many planets discoverable in high-magnification events are not published because of potential degeneracy with stellar mass binaries; alternatively, planets with large $|s-1/s|$ may be missed because follow-up observations tend to only cover the higher-magnification portions of lightcurves. Finally, an inverse power-law with mass ratio $q^{{\lesssim}-0.5}$ could explain the differences between the data and the model, but this seems implausible because we already account for the measured planetary mass function of \citet{Cassan2012}, and this would require that the mass function be roughly twice as steep. Furthermore, a slope twice as steep as that of Cassan et al. (2012) is more than $3$-sigma inconsistent with the constraints on the slope of the planetary mass function reported by \citet{Clanton2016}, derived from a joint analysis of results from microlensing, radial velocity, and direct imaging surveys. Finally, we also tried increasing the $\Delta\chi^2$ threshold for the model and found no significant change in the test statistics until small number statistics in the model became an issue.

\subsection{Problems with the distance estimates?}\label{dataprob}

It is also possible that the distance estimates we use are the cause of the disagreement between the data and the model. As mentioned in \autoref{distmeasure}, the authors that compute Bayesian estimates of the distance and mass would caution us not to use them for statistical studies. However, we should also cast a critical eye upon the parallax measurements that are generally thought to be accurate.

The constraints that shape the posterior distribution of the Bayesian distance estimate come primarily from the measurements of the angular Einstein radius and proper motion that are routine for most planetary microlensing events.\footnote{Analyses that use $\tein$ constraints rather than $\murel$ are using the same information because $\murel=\thetae/\tein$.} $\thetae$ and $\murel$ are derived from measurements of the source radius crossing time $t_{\ast}$, $\tein$ and the source color. $t_{\ast}$ can essentially be ``read off'' from the duration of finite source effects in the lightcurve, and though there is some sensitivity to the choice of limb darkening parameters, it should be determined very robustly given sufficient coverage over the finite source features. $\tein$ and the source color can be smoothly degenerate with blending, which means that uncertainties should account for this degeneracy; however, discrete degeneracies between different models in the $q$--$s$ plane can cause catastrophic errors in $\murel$ and $\thetae$. Bayesian distance estimates can incorporate the multimodal posteriors that such a discrete degeneracy would produce, but our analysis can not. Some events in the observed sample are affected by such a discrete degeneracy \citep[e.g., MOA-2007-BLG-192][]{Bennett2008}, but they are in the minority.

According to our model, the distribution of distances as a function of $\thetae$ is monotonic, and roughly linear in $\log(\thetae/\text{mas})$ with a slope of ${\sim}10$~kpc~dex$^{-1}$ and an r.m.s. scatter of ${\sim}1.3$~kpc in each $0.1$~dex bin (see also \autoref{muthdl} below); the dominant contributor to this scatter is the host mass distribution. Other Galactic models used as priors in Bayesian distance estimates will likely have similar properties. Often $\thetae$ measurements are limited by a systematic error in the conversion of the source color to an angular source size~\citep[e.g.,][]{Yoo2004} with a magnitude of ${\sim}0.05$--$0.1$ in $\ln(\thetae/\text{mas})$. This implies that with only a $\thetae$ measurement, it should be possible to localize the distance of most lenses to within ${\sim}2$~kpc or less. Therefore, we conclude that while the Bayesian estimates of mass are likely to be unreliable, and are likely to conform to our prior on the mass, the Bayesian distance estimates are likely to be much more useful. 

The precision of the Bayesian distance estimates is certainly enough to conduct the present study. Even if one were suspicious of the exact form of the distance distribution that Bayesian estimates use as priors, with such uncertainties a more crude test asking if the number of hosts with distances less than $5$~kpc (beyond which bulge hosts should begin to contribute) is consistent with the model assuming binomial statistics, should be robust. Performing this test, the model predicts $\uwmnltfive\pm\erruwmnltfive$ hosts closer than $5$~kpc, and the model weighted by the \citet{Johnson2010} mass and metallicity abundance trends predicts $\jwmnltfive\pm\errjwmnltfive$ hosts. There are $\nltfive$ hosts closer than $5$~kpc in our sample, implying that our fiducial and mass-metallicity trend weighted models are inconsistent with the data at $\uinconsist$-$\sigma$ and $\jinconsist$-$\sigma$ respectively.

The final ingredient to examine is the parallax distance measurements. As can be seen in \autoref{distdist}, all of the $\npllx$ hosts in events with parallax measurements are within $5$~kpc, which is not surprising given that there is a strong bias toward nearby lenses as discussed in \autoref{distmeasure}. Of the $\npllx$ hosts, $\npllxfour$ are between $4$ and $5$~kpc, $\npllxlttwo$ are closer than $2$~kpc, and $\npllxltone$ are closer than $1$~kpc. By itself, this may not be too surprising given the observational biases. However, if in contrast to the rest of the paper we assume that the fiducial model is accurate, the number of hosts at extremely small distances ($<2$~kpc) is extremely surprising. Given the number of hosts in our sample, our model would predict $\modltone$~hosts in the $0$--$1$~kpc bin and $\modone$ in the $1$--$2$~kpc bin, compared to the three hosts in each (including the Bayesian estimated hosts). In other words, if all the hosts in the first two bins are in the correct bins, the model would predict that the sample would contain $\lots$ planet hosts, twice the number that are actually in the sample! Moving four or five out of the six events with $\dl<2$~kpc to the $6$--$8$~kpc range (where there is the largest deficit of hosts relative to the model), would remove all tension with the model.

So, could the parallax distance estimates be wrong? As described in \autoref{distmeasure}, these require both parallax and finite source measurements, but as discussed earlier in this section, the finite source measurements, which yield $\thetae$ should be fairly robust. Until recently, the only way to measure the microlensing parallax $\pie$ was via orbital parallax.\footnote{Or via terrestrial parallax~\citep{Holz1996, Gould2009}, but this is extremely rare \citep{Gould2013-yee} and was not measured exclusively in any of our sample.} Orbital parallax is an often subtle effect that arises from the departure of the apparent trajectory of the source from linear motion caused by the Earth's acceleration in its orbit. $\pie$ is actually a vector quantity $\vecpie$, with magnitude $\pie=\pirel/\thetae$ and the same direction as the relative proper motion vector. Often only the so-called parallel component of the vector (in the direction parallel to the direction of the Earth's acceleration, which causes an asymmetric distortion to the lightcurve) is well measured, with the perpendicular component (which causes a symmetric distortion) being degenerate with the impact parameter and blending. In addition, orbital parallax is also degenerate with lens orbital motion~\citep{Batista2011,Skowron2011} and source orbital motion~\citep[xallarap, e.g.,][]{Poindexter2005}, which can cause similar apparent changes to the source trajectory. The magnitude of the parallax distortions and their duration (covering the whole magnified potion of the lightcurve) can also make them vulnerable to corruption by long-term systematic trends in photometry.

\begin{figure*}
\includegraphics[width=\textwidth]{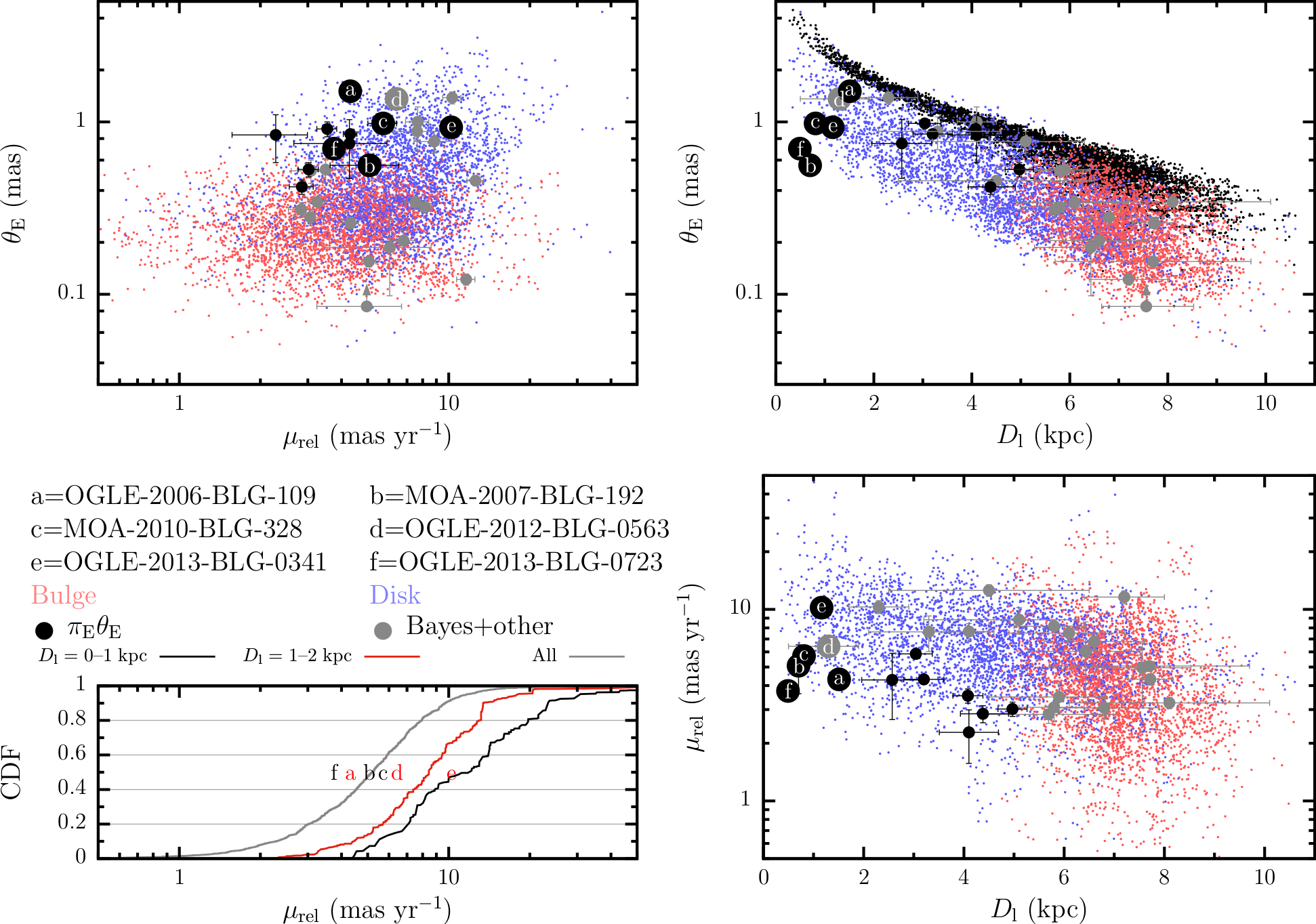}
\caption{\emph{Top row and bottom right panel:} Projections of the $\thetae$-$\murel$-$\dl$ parameter space for a realization of our model (small red points for bulge hosts and blue for disk hosts) and our data sample (large black points for parallax and finite source distance measurements and gray points for Bayesian estimates). Events in the sample with distance estimates $\dl<2$~kpc are individually labeled. In the $\thetae$-$\dl$ panel, small black points show what $\thetae$ would be if all hosts had the same mass ($1\msun$; hosts in the model have masses between $0.08$ and $1\msun$). \emph{Bottom left:} Cumulative distribution of relative proper motions for detected planet hosts in our model for all hosts (gray), hosts between $1$ and $2$~kpc (red) and hosts closer than $1$~kpc (black).}
\label{muthdl}
\end{figure*}

The events with $\dl<2$~kpc are OGLE-2006-BLG-109, MOA-2007-BLG-192, MOA-2010-BLG-328, OGLE-2012-BLG-0563, OGLE-2013-BLG-0341 and OGLE-2013-BLG-0723; we will use abbreviated names (see the notes for \autoref{sample}) for events in our sample from here on. 
We have highlighted them on $\thetae$-$\murel$, $\thetae$-$\dl$ and $\murel$-$\dl$ diagrams in \autoref{muthdl}, which visually compares our host sample to our model. Model points were drawn with replacement from the model, and displaced randomly from their actual values (in order to better visualize point densities) by $\log(\thetae/\text{mas})\pm0.1$, $\log(\murel/\text{mas~yr}^{-1})\pm0.1$ and $\dl\pm0.25$~kpc, where the $\pm$ sign indicates the range of a uniform distribution.

Recalling that $\thetae$ and $\murel$ should be relatively robustly measured, in the $\murel$-$\thetae$ panel of \autoref{muthdl} the nearby hosts do not look to be especially unusual; they cluster toward larger $\thetae$, but are generally within the cloud of model points, with O06-109 (labeled a) being the most extreme. However, in the $\dl$-$\thetae$ and $\dl$-$\murel$ plots, at least three of the nearby hosts (b, c and f, or M07-192, M10-328 and O13-0723, respectively) stand out from the cloud of model points; these three hosts are the ones within $1$~kpc. O06-109 lies at the edge of the model cloud in the $\dl$-$\murel$ plot and O13-0341 lie at the edge in the $\dl$-$\thetae$ plot. O12-0563, being a Bayesian distance estimate, unsurprisingly lies within the cloud of points.

The majority of the scatter in $\thetae$ in the model is caused by the range of lens masses. This can be seen by comparing the $\thetae$ scatter in the $\dl$-$\thetae$ plot between the fiducial red and blue points and the black black points which are the same realization with $\thetae$ plotted as if every star had the same mass of $1\msun$. From this it can easily be seen that the small distance estimates imply extremely small host mass estimates. The host masses in our model range from $0.08$ to $1\msun$. The authors of papers on M07-192, M10-328 and O13-0723 estimate host masses of $0.084$, $0.11$ and $0.031\msun$, respectively~\citep{Kubas2012, Furusawa2013, Udalski2015-venus}. These events have relatively normal angular Einstein radii and relative proper motions, so the low mass and distance estimates are being driven entirely by the parallax measurements. Given that at least some of the events closer than $2$~kpc occupy a region of parameter space that our model predicts should be empty, it is worth looking at each one individually in more detail to see if we can plausibly move them to a larger distance and bring them in line with our model's expectations.

\emph{O06-109} (a) was an extremely complex event with five distinct features from two planets~\citep{Gaudi2008}. Parallax and finite source effects combined to give mass and distance estimates of $0.51\msun$ and $1.51$~kpc. The event also had observable terrestrial parallax~\citep[with a $\Delta\chi^2{\sim}12$][]{Bennett2010}, and $H=17.2$ magnitudes of blended light coincident with the source in AO imaging matches essentially exactly that which is expected from the parallax solution. Given that terrestrial parallax can not be mimicked by xallarap or orbital motion and the bright and unmistakable blended light, we conclude that there is very little chance that the O06-109 distance estimate is significantly in error.
  
\emph{M07-192} (b) was a high-magnification event with poor coverage over the event peak, and only four data points cover the planetary deviation. As such, there were many models (16) that could fit the data well, but these all had similar planetary mass ratios, with the only significant difference being two classes of models with $\thetae$ differing by a factor of ${\sim}2$~\citep{Bennett2008}; even if the solution with the smaller $\thetae$ were chosen, the distance to the host would be less than $2$~kpc, but the relative proper motion would be reduced to $2$--$3$~mas~yr$^{-1}$. \citet{Bennett2008} found xallarap models that fit the data, but argued that these had a low probability relative to the parallax model. From AO imaging, \citet{Kubas2012} find that there is a $K_{\mathrm{s}}=19.2$ star blended with the source, but do not measure its color very precisely. \citet{Kubas2012} do not question the parallax measurement of \citet{Bennett2008} so interpret the blended light as being due to a $0.08$-$\msun$ star nearby, but the $K_{\mathrm{s}}$ magnitude they measure is compatible with a $0.7$-$\msun$ star at ${\sim}7$~kpc, and would only cause some tension with their $0.37$~mag $J-K_{\mathrm{s}}$ error bar. \citet{Bennett2008} estimate that the a priori probability of the parallax signal being caused by xallarap is $0.32$~percent, accounting for the distribution of binary periods, but not accounting for the increased stellar density in the bulge relative to the nearby disk. They guess that the increased density in the bulge would increase this probability by a factor of about $7$, but do not rigorously compute this number. Given the difficulty in computing such a probability, it might be reasonable to assume that it might actually be significantly larger. 

\emph{M10-328} (c) was a low-magnification event with a ${\sim}5$~day long planetary deviation that was poorly covered due to bad weather, except for the caustic exit over a cusp~\citep{Furusawa2013}. The authors find a xallarap model that fits better than the parallax model by $\Delta\chi^2\sim 5$, but argue that the parallax interpretation is most likely because an unconstrained xallarap fit with a period fixed to $1$~yr yields an orbital orientation that is coincident with the Earth's relative to the event line of sight. We agree that this would seem to rule out xallarap as an impostor for parallax. Additionally, the large angular Einstein radius of the event would imply that if the host were in the bulge, it would be ${\sim}1\msun$, and therefore bright unless it were a stellar remnant. It therefore seems very unlikely that we could plausibly place this host in the bulge. The only remaining hope of doing so is that the poor coverage of the planetary anomaly and the combination of a cusp crossing with orbital motion on the single caustic exit result in a larger than reported uncertainty on the angular Einstein radius or a misestimation of it.

\emph{O12-0563} (d) was a high-magnification event that was densely covered over the peak~\citep{Fukui2015}. Surprisingly, despite the source not crossing the central caustic, the authors claim they are able to measure $\thetae$ to be $1.36$~mas to a precision of ${\sim}10$~percent. While formally the authors measure parallax, they argue that the detection is likely due to systematics, and so we do not count it among the events with parallax detections. AO imaging shows blended light of $K_{\mathrm{s}}=17.7$. The authors do not consider a xallarap model. The large $\thetae=1.36$~mas of the event would by itself exclude the possibility of a bulge lens, unless it were a neutron star or black hole. Even if $\thetae$ were overestimated, it would need to be severely overestimated to allow us to move the lens into the bulge.

\emph{O13-0341} (e) is a triple lens event with a very small planetary signal (the dip) that by itself might be overlooked~\citep{Gould2014}. However, the well-covered crossing of a large, binary-star central-caustic crossing is also better fit if a planet with the same properties as would cause the dip is included in the model. A low amplitude bump during an earlier observing season identifies the binary as a wide binary, and so the planet orbits one of the stars. The parallax signal in the lightcurve is strong. The authors do not mention if they considered a xallarap model, but the large proper motion and large angular Einstein radius place the event in a region of parameter space that is almost entirely occupied by disk stars. 

\emph{O13-0723} (f) is a second triple lens planetary event discovered in 2013. It also seems to have a strong parallax signal~\citep{Udalski2015-venus}. Again, the planet orbits only one of the binary stars, however in this case the argument for this scenario over a circumbinary planet is based upon an argument that the parameters of the circumbinary model require fine tuning. The authors do not mention if they considered a xallarap model, but unlike O13-0341, the location of the event in the $\murel$-$\thetae$ plane is occupied by both bulge and disk stars and such a model should be considered.\footnote{Note that since submission, an alternative, two-body lens model with orbital motion has been found for O13-0723, with lower $\chi^2$ and a parallax that implies a distance six times further away~\citep{Han2016}. However, this new model is not without problems, as during a short anomaly, it contains an even shorter caustic crossing in a data gap whose entry begins immediately after data taking ends. This can occur if a model search routine is trapped in a local rather than a global minimum.}

While both O13-0341 and O13-0723 seem to have strong parallax signals, it is unsettling to say the least that the only two planets discovered in binary star microlensing events with caustic crossings also have extremely small host distance (and mass) estimates.\footnote{There is one other planetary event in a binary system, O08-092~\citep{Poleski2014}, but the event was low-magnification and none of the caustics associated with the binary star was crossed.} There have been other binary microlensing events with small mass and distance estimates, e.g., OGLE-2009-BLG-151/MOA-2009-BLG-232 ($0.39$~kpc, $0.025\msun$ total mass), OGLE-2011-BLG-0420 \citep[$1.99$~kpc, $0.034\msun$][]{Choi2013}, OGLE-2012-BLG-0358~\citep[$1.73$~kpc, $0.024\msun$][]{Han2013}, OGLE-2013-BLG-0578~\citep[$1.16$~kpc, $0.16\msun$][]{Park2015} and MOA-2011-BLG-149~\citep[$1.07$~kpc, $0.16\msun$][]{Shin2012}. For these binaries it is easy to dismiss the unusually small masses and distances as a publication bias: were they further away, parallax would not be measurable, which would mean their low-mass would go unnoticed, and there would be nothing interesting to write a paper about them. However, this bias is less prominent for planetary microlensing events as all planets are publishable, though there could be a delay in publishing ``less interesting'' ones without parallax measurements, which would bias planet in binary detections to smaller distances. It is perhaps more likely that planets in binaries would suffer a bias that works in the opposite direction, due to the possibility that parallax effects in addition to the triple lens nature, would make the events more difficult to model and hence delay publication. 

One might think that there is another possible bias that could result in us only detecting planets in binary systems when they are nearby. That is, if the binary star prevents the formation or survival of planets on wider orbits, then it will only be possible to detect the surviving, closer-in planets in events with small physical Einstein ring radii $\re$. However, $\re\propto\sqrt{\dl(\ds-\dl)}$, and so there are lenses with small physical Einstein ring radii near their sources in the bulge as well as near us the observer; there should be many more planets discovered in binary systems in the bulge than in the nearby disk. We are therefore forced to ask the question: is there something going wrong with the modeling of parallax in triple lens microlensing events (and maybe in binary star events too)? Given the complexity of triple lens models it seems plausible that something could be going wrong, but it is difficult to guess at what this could be, especially in the case of O13-0341 where the parallax, angular Einstein radius and relative proper motion seem to tell a consistent story. 

It is possible to test the predictions of the parallax models to some degree. They predict the direction of the relative source-lens proper motion, and after ${\sim}10$ years the lens and source may separate enough for this direction to be measured. An expedited method to test the validity of the parallax models would be to model only the ground-based data for upcoming binary microlensing events that are observed by Spitzer~\citep[e.g.,][]{Zhu2015-binary, Udalski2015-spitzer} or {\it K2} Campaign 9~\citep{Howell2014,Henderson2015}, and then test the orbital parallax prediction with the much more robust Spitzer or {\it K2} satellite parallax measurements. \citet{Udalski2015-spitzer} performed this test for a planetary event and found that the orbital and Spitzer-measured satellite parallaxes agreed, but the binary event modeled by \citet{Zhu2015-binary} did not have a detectable orbital parallax signature, and alas Spitzer did not observe the past events for which we are worried about their parallax solutions.

While we can offer no convincing evidence to do so, potentially two of the six hosts with $\dl<2$~kpc (M07-192 and O13-0723) could plausibly be moved to larger distances if xallarap is mimicking parallax in either event or if there were errors in the parallax modeling of O13-0723. It is worth noting that \citet{Poindexter2005} found that 5 out of 22 single lens events with parallax-like signals that they modeled were more likely to have been caused by xallarap. We might therefore expect that ${\sim}20$~percent of our $\npllx$ parallax events (i.e., ${\sim}2$--$3$) would actually be caused by xallarap. In fact, O07-368 should also be added to our parallax-like sample, because it has a reasonable parallax solution, but \citet{Sumi2010} find that it is better fit and more likely caused by xallarap. If the parallax solution of O07-368 had been believed blindly without considering the possibility of xallarap, the event would also have been in our sub-sample with suspiciously small distance estimates less than $2$~kpc. If we remove M07-192 and O13-0723 from our sample, AD tests like those we performed in \autoref{paucity} have $p$-values greater than $0.01$ for $\fbulge<\fbmaxonesusp$, and so would go most of the way toward explaining the discrepancy between the model and the observed data.

\subsection{Combining Corrections}

\begin{figure}
\includegraphics[width=\columnwidth]{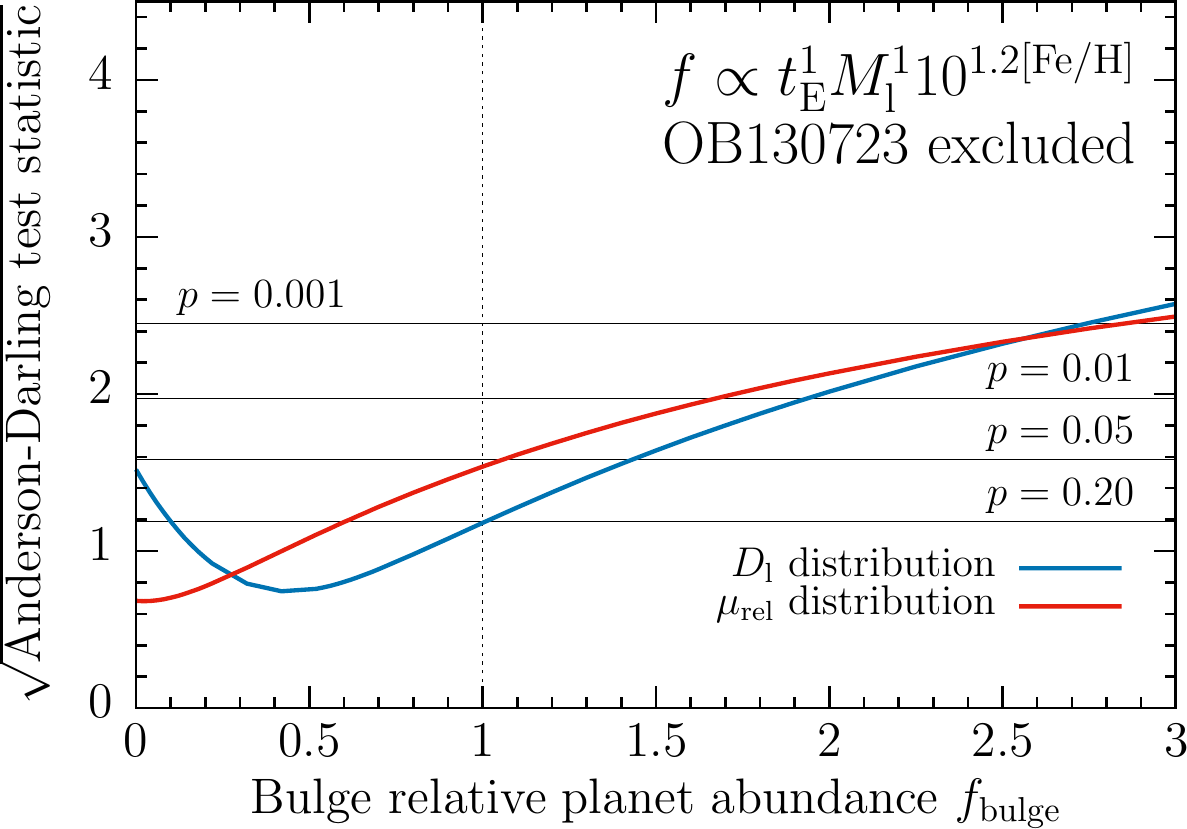}
\caption{AD test statistics as a function of $\fbulge$ for the lens distance (blue) and relative proper motion (red) distributions, after removing OB130723 and
weighting the model by $\tein^{+1}$ (to match the timescale distribution), $M^{+1}$ and $10^{+1.2\text{[Fe/H]}}$ (to match radial velocity planet abundance trends).}
\label{amalgamated}
\end{figure}

We have found that various factors can help to bring the model distribution towards consistency with the null hypothesis, namely that the planet abundance in the bulge is no different to that in the disk, but do not go all the way. We now investigate whether combining all the factors can make the model consistent. We assume that planet abundance is a function of both host mass and metallicity with the same dependence as found by radial velocity surveys \citep[$\propto M 10^{1.2\text{[Fe/H]}}$]{Johnson2010}. We also assume that the \citetalias{Henderson2014-kmt} simulations do not fully capture the increased sensitivity to planets in longer timescale events for planets detected by follow-up observations, and so weight the simulated planet detections by $\tein^{1}$. Finally, we remove the event OB130723 from our sample of distance estimates, due to a non-planetary model being found that better explains the data on the event~\citep{Han2016}, which was reported after this paper was submitted.

\autoref{amalgamated} shows the AD statistic as a function of $\fbulge$ for the model with all of the combined corrections, for both the distance and relative proper motion distributions. For $\fbulge=1$, the model is much more consistent with the both the $\dl$ and $\murel$ distributions, with $p>0.05$ in both cases. The full range of $\fbulge$ allowed with a threshold of $p>0.01$ for both $\dl$ and $\murel$ distributions is $\fbulge\lesssim 1.7$. This might suggest that the puzzle has been solved. However, there is still some tension with the $\fbulge=1$ scenario, with significantly smaller values of $\fbulge$ providing better fits to the $\dl$ and $\murel$ distributions. We expect that further events with erroneous parallax measurements implying very small distance estimates are likely to be the cause of this remaining tension.

\section{Conclusions}\label{conclusion}

Using the Anderson-Darling test, we have found that the distribution of distance estimates for the hosts of planets discovered in microlensing surveys is inconsistent with the distribution expected from a simulation of a planetary microlensing survey that incorporates a reasonable Galactic model. The model can be brought into consistency with the observed distribution if the relative abundance of planets in the Galactic bulge is reduced relative to that of the disk by a factor $\fbulge\le\fbmaxone$ at $99$~percent confidence (model dependent).

Does this mean that the Galactic bulge devoid of planets? We can not completely rule out this conclusion, but there are a number of lines of reasoning that suggest it is unlikely. Firstly, the relative proper motion distribution of the hosts would be in tension with a model that was completely devoid of bulge planets. Secondly, incorporation of a plausible mass and metallicity dependence to the planet abundance increases the upper limit on $\fbulge$ to $\jfbmaxone$. Thirdly, the inclusion of an additional dependence of planet detection efficiency on some microlensing parameters that is not captured by our model can increase the consistency of the model with the data, especially the event timescale. Finally, some of the inconsistency of the model with the data might be removed if a relatively small set of the distance estimates are wrong. Specifically, we suspect that maybe one third of events with parallax distance estimates smaller than $2$~kpc might actually have other signals such as xallarap mimicking parallax signals and thus causing completely erroneous distance estimates. A recent paper by \citet{Han2016} has confirmed our suspicions for one event.

Our study has shown that in principle it is possible to infer relative properties of the Galactic distribution of planets (such as the relative abundance of planets in the disk versus the bulge) without the need to compute the absolute abundances of planets in each population as advocated by \citet{CalchiNovati2015,Yee2015} and \citet{Zhu2015-planetsens}. Studies similar to ours will be complimentary to the space-based parallax programs that are now underway with Spitzer and are forthcoming with {\it K2}~\citep{Howell2014, Henderson2015}. The space-based parallax campaigns will enable precise parallax distance measurements for almost all planetary microlensing events they observe, but these will be limited to relatively small samples of planets. We have pioneered an approach that while model dependent, can be used with large samples of microlensing planets that will be found by second-generation microlensing planet searches such as OGLE-IV~\citep{Udalski2015-ogleiv}, MOA-II~\citep{Sako2007} and KMTNet~\citepalias{KMTref, Henderson2014-kmt} that will have selection effects that are much easier to model than those of our current sample. Such samples are beginning to be assembled (\citeauthor{Shvartzvald2016}~\citeyear{Shvartzvald2016}; Suzuki et al., in prep.). Finally, the massive samples of planets that will be found by Euclid~\citep{Penny2013} and WFIRST~\citep{Spergel2015} it should be possible to probe the distribution of various planet populations (e.g., Jupiters compared to Earths) as a function of their host's location within the Galaxy.

\vspace{12pt}
We would like to thank Radek Poleski, Wei Zhu, Andy Gould, Dave Bennett and the referee for their suggestions. Work by MTP was performed in part under contract with the California Institute of Technology (Caltech)/Jet Propulsion Laboratory (JPL) funded by NASA through the Sagan Fellowship Program executed by the NASA Exoplanet Science Institute. Work by CBH was supported by an appointment to the NASA Postdoctoral Program at the Jet Propulsion Laboratory, administered by Oak Ridge Associated Universities through a contract with NASA.

\bibliographystyle{aasjournal}
\bibliography{libraryshort,apj-jour}

\end{document}

%% file: defs.tex
\newcommand{\thetae}{\theta_{\mathrm{E}}}
\newcommand{\re}{r_{\mathrm{E}}}
\newcommand{\pie}{\pi_{\mathrm{E}}}
\newcommand{\vecpie}{\vec{\pi}_{\mathrm{E}}}
\newcommand{\dl}{D_{\mathrm{l}}}
\newcommand{\ds}{D_{\mathrm{s}}}

\newcommand{\tein}{t_{\mathrm{E}}}
\newcommand{\pirel}{\pi_{\mathrm{rel}}}
\newcommand{\murel}{\mu_{\mathrm{rel}}}
\newcommand{\murelgeo}{\mu_{\mathrm{rel},\earth}}
\newcommand{\murelhelio}{\mu_{\mathrm{rel},\sun}}

\newcommand{\uzero}{u_{\mathrm{0}}}

\newcommand{\msun}{M_{\odot}}
\newcommand{\mearth}{M_{\earth}}

\newcommand{\dd}{\mathrm{d}}
\newcommand{\mpl}{M_{\mathrm{p}}}

%% file: galdistmacros.tex
\newcommand{\nplanets}{31}
\newcommand{\admodel}{6.661}
\newcommand{\padmodel}{5.0\times10^{-4}}
\newcommand{\addisk}{1.468}
\newcommand{\paddisk}{0.18}

\newcommand{\addiskmu}{2.704}
\newcommand{\paddiskmu}{0.039}
\newcommand{\minal}{0.9}
\newcommand{\minalad}{5.263}
\newcommand{\pminalad}{2.2\times10^{-3}}
\newcommand{\alminmu}{-1}
\newcommand{\fbmin}{0.02}
\newcommand{\fbmax}{0.32}
\newcommand{\fbmaxone}{0.54}
\newcommand{\fbmaxonesusp}{0.96}
\newcommand{\jfbmin}{0.06}
\newcommand{\jfbmax}{0.32}
\newcommand{\jfbmaxone}{0.7}
\newcommand{\presample}{1.7\times10^{-3}}
\newcommand{\ngepone}{3}
\newcommand{\npllx}{12}
\newcommand{\npllxfour}{4}
\newcommand{\npllxlttwo}{5}
\newcommand{\npllxltone}{3}
\newcommand{\modltone}{0.26}
\newcommand{\modone}{1.24}
\newcommand{\lots}{61.9}
\newcommand{\uwmnltfive}{9.0}
\newcommand{\erruwmnltfive}{2.5}
\newcommand{\jwmnltfive}{9.4}
\newcommand{\errjwmnltfive}{2.6}
\newcommand{\nltfive}{17}
\newcommand{\uinconsist}{3.2}
\newcommand{\jinconsist}{3.0}

%% file: compacttable.tex
\begin{tabularx}{\textwidth}{lXXXXXXXl}
\hline
Event & $D_{\mathrm{l}}$ & $\mu_{\mathrm{rel}}$ & $\thetae$ & $\log q$ & $\ell$ & $b$ & Method & Reference \\
      & (kpc)    & (mas~yr$^{-1}$) & (mas) & & (\degr) & (\degr) & & \\
\hline
O03-235 & $5.8_{-0.7}^{+0.6}$ &$3.1\pm 0.5$ & $0.52\pm 0.08$ & $-2.4$ &$+2.20$ & $-3.69$ & Bayes & 1,2\\
O05-071 & $3.2\pm0.4$ & $4.3\pm 0.2$ & $0.85\pm 0.05$ & $-2.1$ &$-4.42$ & $-3.79$ & $\pi_{\mathrm{E}}\theta_{\mathrm{E}}$ & 3,4\\
O05-169 & $4.1\pm0.4$ & $7.7\pm 0.8$ & $1.00\pm 0.22$ & $-4.2$ &$+0.68$ & $-4.74$ & Bayes & 5,6,7\\
O05-390 & $6.6\pm1.0$ & $6.8\pm 0.7$ & $0.20\pm 0.03$ & $-4.1$ &$-0.27$ & $-2.36$ & Bayes & 8\\
O06-109 & $1.5\pm0.1$ & $4.3\pm 0.1$ & $1.50\pm 0.04$ & $-3.3$ &$-0.21$ & $-1.89$ & $\pi_{\mathrm{E}}\theta_{\mathrm{E}}$ & 9,10\\
M07-192 & $0.70_{-0.12}^{+0.21}$ &$5.1\pm 1.4$ & $0.56\pm 0.06$ & $-4.4$ &$+4.03$ & $-3.39$ & $\pi_{\mathrm{E}}\theta_{\mathrm{E}}$ & 11,12,13\\
O07-368 & $5.9_{-1.4}^{+0.9}$ &$3.5\pm 0.6$ & $0.53\pm 0.08$ & $-4.0$ &$-1.65$ & $-3.69$ & Bayes & 14\\
M07-400 & $5.8_{-0.8}^{+0.6}$ &$8.2\pm 0.5$ & $0.32\pm 0.02$ & $-2.6$ &$+2.38$ & $-4.70$ & Bayes & 15\\
O08-092 & $8.1\pm2.0$ & $3.2\pm 0.2$ & $0.34\pm 0.02$ & $-3.6$ &$-4.75$ & $-3.34$ & other & 16\\
M08-310 & $7.7\pm2.0$ & $5.1\pm 0.3$ & $0.15\pm 0.01$ & $-3.5$ &$-4.09$ & $-4.56$ & Bayes & 17\\
O08-355 & $6.8\pm1.1$ & $3.1\pm 0.4$ & $0.28\pm 0.03$ & $-1.9$ &$-0.08$ & $-3.45$ & Bayes & 18\\
M08-379 & $3.3_{-1.2}^{+1.3}$ &$7.6\pm 1.6$ & $0.88\pm 0.19$ & $-2.2$ &$+0.38$ & $-3.11$ & Bayes & 19\\
M09-266 & $3.0\pm0.3$ & $5.9\pm 0.3$ & $0.98\pm 0.04$ & $-4.2$ &$-4.93$ & $-3.58$ & $\pi_{\mathrm{E}}\theta_{\mathrm{E}}$ & 20\\
M09-319 & $6.1_{-1.2}^{+1.1}$ &$7.5\pm 0.7$ & $0.34\pm 0.03$ & $-3.4$ &$+4.20$ & $-3.01$ & Bayes & 21\\
M09-387 & $5.7\pm1.3$ & $2.8\pm 0.3$ & $0.31\pm 0.03$ & $-1.9$ &$-3.44$ & $-4.09$ & Bayes & 22\\
M10-328 & $0.81\pm0.10$ & $5.7\pm 0.7$ & $0.98\pm 0.12$ & $-3.6$ &$-0.16$ & $-3.21$ & $\pi_{\mathrm{E}}\theta_{\mathrm{E}}$ & 23\\
M10-353 & $6.4\pm1.1$ & $6.0\pm 2.6$ & $0.19\pm 0.09$ & $-2.9$ &$+3.60$ & $-2.90$ & Bayes & 24\\
M10-477 & $2.3\pm0.6$ & $10.3\pm 0.8$ & $1.38\pm 0.11$ & $-2.7$ &$+0.05$ & $-5.09$ & Bayes & 25\\
O11-0251 & $2.6\pm0.6$ & $4.3\pm 1.6$ & $0.75\pm 0.28$ & $-2.7$ &$+0.67$ & $+2.33$ & $\pi_{\mathrm{E}}\theta_{\mathrm{E}}$ & 26\\
M11-262 & $7.2\pm0.8$ & $11.6\pm 0.9$ & $0.12\pm 0.01$ & $-3.4$ &$-0.37$ & $-3.92$ & Bayes & 27\\
O11-265 & $4.4\pm0.5$ & $2.9\pm 0.3$ & $0.42\pm 0.04$ & $-2.4$ &$+2.70$ & $-1.52$ & $\pi_{\mathrm{E}}\theta_{\mathrm{E}}$ & 28\\
M11-293 & $7.7\pm0.4$ & $4.3\pm 0.3$ & $0.26\pm 0.02$ & $-2.3$ &$+1.52$ & $-1.65$ & Bayes & 29,30\\
M11-322 & $7.6_{-0.9}^{+1.0}$ &$5.0\pm 1.7$ & $>0.085$ &$-1.5$ &$+3.63$ & $-2.81$ & Bayes & 31\\
O12-0026 & $4.1\pm0.3$ & $3.5\pm 0.3$ & $0.91\pm 0.09$ & $-3.9$ &$+0.19$ & $+3.06$ & $\pi_{\mathrm{E}}\theta_{\mathrm{E}}$ & 32\\
O12-0406 & $5.0\pm0.3$ & $3.0\pm 0.3$ & $0.53\pm 0.05$ & $-2.2$ &$-0.46$ & $-2.22$ & $\pi_{\mathrm{E}}\theta_{\mathrm{E}}$ & 33,34\\
O12-0563 & $1.3_{-0.8}^{+0.6}$ &$6.4\pm 0.6$ & $1.36\pm 0.13$ & $-3.0$ &$+3.31$ & $-3.25$ & Bayes & 35\\
M13-220 & $4.5\pm2.0$ & $12.6\pm 0.9$ & $0.46\pm 0.03$ & $-2.5$ &$+1.50$ & $-3.76$ & other & 36\\
O13-0341 & $1.2\pm0.1$ & $10.2\pm 0.8$ & $0.93\pm 0.07$ & $-4.3$ &$-0.05$ & $-1.68$ & $\pi_{\mathrm{E}}\theta_{\mathrm{E}}$ & 37\\
O13-0723 & $0.49\pm0.04$ & $3.8\pm 0.3$ & $0.70\pm 0.06$ & $-4.2$ &$-0.02$ & $+2.83$ & $\pi_{\mathrm{E}}\theta_{\mathrm{E}}$ & 38\\
O14-0124 & $4.1\pm0.6$ & $2.3\pm 0.7$ & $0.84\pm 0.26$ & $-3.2$ &$+2.34$ & $-2.92$ & $\pi_{\mathrm{E}}\theta_{\mathrm{E}}$ & 39\\
MOA-bin-1 & $5.1_{-1.9}^{+1.2}$ &$8.8\pm 1.4$ & $0.77\pm 0.11$ & $-2.3$ &$-0.11$ & $-1.48$ & Bayes & 40\\
\hline
\end{tabularx}\\
\textbf{Notes.} Lens distance estimates, proper motions, mass ratios and galactic coordinates for our sample of microlensing planet hosts. The method column is explained in the text. Event names of the format SURVEY-YYYY-BLG-NNN(N) are abbreviated as SYY-NNN(N) with O$=$OGLE and M$=$MOA, e.g., OGLE-2014-BLG-0124 becomes O14-0124. Events with joint survey names are listed only by the first name.\\
\textbf{References.} (1) \citealt{Bennett2006}; (2) \citealt{Bond2004}; (3) \citealt{Dong2009a}; (4) \citealt{Udalski2005}; (5) \citealt{Bennett2015}; (6) \citealt{Batista2015}; (7) \citealt{Gould2006}; (8) \citealt{Beaulieu2006}; (9) \citealt{Bennett2010}; (10) \citealt{Gaudi2008}; (11) \citealt{Kubas2012}; (12) \citealt{Gould2010a}; (13) \citealt{Bennett2008}; (14) \citealt{Sumi2010}; (15) \citealt{Dong2009}; (16) \citealt{Poleski2014}; (17) \citealt{Janczak2010}; (18) \citealt{Koshimoto2014}; (19) \citealt{Suzuki2014}; (20) \citealt{Muraki2011}; (21) \citealt{Miyake2011}; (22) \citealt{Batista2011}; (23) \citealt{Furusawa2013}; (24) \citealt{Rattenbury2015}; (25) \citealt{Bachelet2012}; (26) \citealt{Kains2013}; (27) \citealt{Bennett2014}; (28) \citealt{Skowron2015}; (29) \citealt{Batista2014}; (30) \citealt{Yee2012}; (31) \citealt{Shvartzvald2014}; (32) \citealt{Han2013a}; (33) \citealt{Tsapras2014}; (34) \citealt{Poleski2014}; (35) \citealt{Fukui2015}; (36) \citealt{Yee2014}; (37) \citealt{Gould2014}; (38) \citealt{Udalski2015-venus}; (39) \citealt{Udalski2015-spitzer}; (40) \citealt{Bennett2012}